\documentclass[aps,prd,preprintnumbers,superscriptaddress,nofootinbib,showpacs]{revtex4}
\usepackage[dvips]{graphicx}
\usepackage{bm,latexsym,amsmath,amssymb,amsfonts,mathrsfs}
\usepackage{color}
\input{colordvi.tex}
\newcommand*{\D}{{\rm d}}

\begin{document}

\title{Multi-field G-inflation}

\author{Tsutomu~Kobayashi}
\email[Email: ]{tsutomu"at"rikkyo.ac.jp}
\affiliation{Department of Physics, Rikkyo University, Toshima, Tokyo 175-8501, Japan
}

\author{Norihiro~Tanahashi}
\email[Email: ]{norihiro.tanahashi"at"ipmu.jp}
\affiliation{Kavli Institute for the Physics and Mathematics of the Universe (WPI), 
The University of Tokyo, Kashiwa, Chiba 277-8583, Japan
}

\author{Masahide~Yamaguchi}
\email[Email: ]{gucci"at"phys.titech.ac.jp}
\affiliation{Department of Physics, Tokyo Institute of Technology, Tokyo 152-8551, Japan
}

\begin{abstract}
We propose a multi-field extension of (generalized) G-inflation,
based on covariant multi-galileons and their generalization preserving
second-order field equations. We compute the quadratic action for cosmological
perturbations. By comparing the formulas for cosmological perturbations,
it is highlighted that multi-field DBI galileon inflation is not included
in the multi-field version of generalized G-inflation.
Our result indicates that the generalized covariant multi-galileon theory
is {\em not} the most general multi-scalar-tensor theory with
second-order field equations.
\end{abstract}

\preprint{RUP-13-7, IPMU13-0158}
\maketitle

\section{Introduction}

Inflation is now the basis of modern cosmology, supported
phenomenologically and observationally.  The precise measurements of the
cosmic microwave background temperature anisotropies made by
WMAP~\cite{Bennett:2012zja,Hinshaw:2012aka} and
Planck~\cite{Ade:2013ktc,Ade:2013zuv,Ade:2013uln} enable us to probe the
nature of quantum fluctuations generated during inflation.  Further
progress in theory and observations will help our quest for the
origins of inflation and large scale structure of the Universe.

To constrain existing inflation models and hunt for {\em the}
inflaton(s) with cosmological observations, it is important to develop a
theoretical framework to address the most general inflation models
possible.  Within a single-field class, a powerful framework
``generalized G-inflation'' was proposed~\cite{Kobayashi:2011nu}, in
which all the single-field inflation models are described by Horndeski's
most general scalar-tensor Lagrangian~\cite{Horndeski,
Charmousis:2011bf} in a unified manner.  The generalized G-inflation
framework generates novel models as well, such as
G-inflation~\cite{Kobayashi:2010cm, Deffayet:2010qz} and Higgs
G-inflation~\cite{Kamada:2010qe}.  Aspects of cosmological perturbations
from generalized G-inflation have been studied extensively in
Refs.~\cite{Mizuno:2010ag, Naruko:2011zk, Kobayashi:2011pc, Gao:2011mz,
Gao:2011qe, RenauxPetel:2011sb, DeFelice:2011uc, Gao:2011vs, Gao:2012ib,
Takamizu:2013gy, Frusciante:2013haa, DeFelice:2013ar}.  Planck
constraints on generic single-field inflation of Horndeski's class are
summarized in Ref.~\cite{Tsujikawa:2013ila}.

In this paper, we go beyond the single-field class and consider a
multi-field extension of generalized G-inflation.  A variety of
multi-field inflation models have been proposed and their primordial
perturbations have been calculated so far, including multi-DBI
inflation~\cite{Easson:2007dh,Huang:2007hh,Langlois:2008mn,Langlois:2008wt,Langlois:2008qf,Langlois:2009ej,RenauxPetel:2009sj}
and multi-field inflation with nonminimal couplings.
Our aim is to develop a general and compact framework to treat such
multi-field models in a unified manner.  To do so, we utilize the
covariant multi-galileons and their generalization, which were recently
proposed and conjectured to be the most general multi-scalar-tensor
theory with second-order field equations~\cite{Padilla}.  We compute the
quadratic action for cosmological perturbations from the generalized
multi-galileons, and in so doing we investigate whether the conjecture
is true or not.

This paper is organized as follows. In the next section, we briefly review
galileons and their generalization in both the single-field and multi-field cases.
The background equations are
derived in Sec.~III, and the generic second-order actions for cosmological tensor and scalar
perturbations are computed in Sec.~IV.
In Sec.~V, we discuss whether or not the generalized multi-galileon theory
is the most general multi-scalar-tensor theory. The final section is devoted to the conclusion.
In the Appendix, the covariant equations of motion for generalized multi-galileons
are presented for completeness.

\section{Generalizing single- and multi-galileons}

\subsection{Generalized galileon}

Let us start with reviewing briefly how the single-field galileon theory
is generalized to the most general scalar-tensor theory.
See e.g.
Ref.~\cite{Deffayet:2013lga} for a more detailed construction of those galileon theories.

The galileon scalar field was originally considered in Minkowski spacetime~\cite{Nicolis:2008in},
which can be covariantized to incorporate gravity~\cite{Deffayet:2009wt}.
In doing so, nonminimal couplings to
the Ricci scalar $R$ and the Einstein tensor $G_{\mu\nu}$ are inevitably introduced
to maintain second-order field equations for
the scalar field and the metric.
The resultant Lagrangian is given by
\begin{eqnarray}
{\cal L}&=&c_2X-c_{3}X\Box\phi+\frac{c_4}{2}X^2R+c_{4}X\left[
(\Box\phi)^2-(\nabla_\mu\nabla_\nu\phi)^2\right]
\nonumber\\&&
+c_5X^2G^{\mu\nu}\nabla_{\mu}\nabla_\nu\phi-\frac{c_5}{3}X
\left[(\Box\phi)^3-3\Box\phi(\nabla_\mu\nabla_\nu\phi)^2+2(\nabla_\mu\nabla_\nu\phi)^3\right],
\end{eqnarray}
where
$(\nabla_\mu\nabla_\nu\phi)^2:=\nabla_\mu\nabla_\nu\phi\nabla^\mu\nabla^\nu\phi$,
$(\nabla_\mu\nabla_\nu\phi)^3:=\nabla_\mu\nabla_\nu\phi\nabla^\nu\nabla^\lambda\phi
\nabla_\lambda \nabla^\mu\phi$, and
$X:=-(\partial\phi)^2/2$ is the usual kinetic term of the scalar field $\phi$.
The covariant version of the galileon can further be generalized
by promoting $X$ and $X^2$ in the Lagrangian to
arbitrary functions of $\phi$ and $X$,
leading to~\cite{Deffayet:2011gz}
\begin{eqnarray}
{\cal L}&=&G_2(X, \phi)-G_{3}(X, \phi)\Box\phi+G_4(X, \phi)R+\frac{\partial G_{4}}{\partial X}\left[
(\Box\phi)^2-(\nabla_\mu\nabla_\nu\phi)^2\right]
\nonumber\\&&
+G_5(X, \phi)G^{\mu\nu}\nabla_{\mu}\nabla_\nu\phi-\frac{1}{6}\frac{\partial G_{5}}{\partial X}
\left[(\Box\phi)^3-3\Box\phi(\nabla_\mu\nabla_\nu\phi)^2+2(\nabla_\mu\nabla_\nu\phi)^3\right],
\label{GG}
\end{eqnarray}
which still has second-order field equations.  Interestingly, it turns
out that the generalized galileon theory thus constructed in four
dimensions\footnote{The generalized galileon theory can be formulated
also in spacetime dimensions higher than four, but there is no proof
that it is the most general scalar-tensor theory in higher dimensions.}
is in fact the most general one having second-order field equations both
for $\phi$ and the metric~\cite{Kobayashi:2011nu}, {\em i.e.}, the
Horndeski theory developed forty years
ago~\cite{Horndeski}.\footnote{See also Ref.~\cite{disf} for a
recent consideration on the relation between the second-order nature of
the Horndeski theory and the generic disformal transformation of the
metric.}

The nontrivial point is that
by construction it is not guaranteed that the generalized galileon theory
includes the terms that would vanish in the Minkowski limit.
For example,
the following terms give rise to second-order field equations,\footnote{The
term~(\ref{DDRiem}) has been studied e.g.\ in Ref.~\cite{deRham:2011by} in the
context of massive gravity. Thanks to the transverse nature of the double dual Riemann
tensor, $\nabla_\mu L^{\mu\alpha\nu\beta}=0$, the equations of motion
derived from this term remain of second order.}
but could be dropped:
\begin{eqnarray}
&&
\xi(\phi)\left(R_{\mu\nu\rho\sigma}R^{\mu\nu\rho\sigma}-4R_{\mu\nu}R^{\mu\nu}+R^2\right),
\label{NMGB}
\\
&&
L^{\mu\alpha\nu\beta}\partial_\mu\phi\partial_\nu\phi\nabla_\alpha\nabla_\beta\phi,
\label{DDRiem}
\end{eqnarray}
where $L^{\mu\alpha\nu\beta}$ is the double dual Riemann tensor,
\begin{eqnarray}
L^{\mu\alpha\nu\beta}:=
R^{\mu\alpha\nu\beta}+
\left(R^{\mu\beta}g^{\nu\alpha}+R^{\nu\alpha}g^{\mu\beta}
-R^{\mu\nu}g^{\alpha\beta}-R^{\alpha\beta}g^{\mu\nu}\right)
+\frac{1}{2}R\left(g^{\mu\nu}g^{\alpha\beta}-g^{\mu\beta}g^{\nu\alpha}\right).
\label{def-ddriem}
\end{eqnarray}
In fact, however, the nonminimal coupling to the Gauss-Bonnet term~(\ref{NMGB})
can be reproduced from~\cite{Kobayashi:2011nu}
\begin{eqnarray}
G_2=8\xi^{(4)} X^2(3-\ln X),
\quad
G_3=4\xi^{(3)} X(7-3\ln X),
\quad
G_4=4\xi^{(2)} X(2-\ln X),
\quad
G_5=-4\xi^{(1)}\ln X,
\quad
\xi^{(n)}:=\frac{\partial^n\xi}{\partial\phi^n},
\end{eqnarray}
while the term containing the double dual Riemann tensor~(\ref{DDRiem})
simply from $G_5=X$~\cite{Narikawa:2013pjr}.
A lesson drawn from the above fact is that
even though
the Lagrangian~(\ref{GG}) does not depend on the Riemann tensor explicitly,
one can carefully choose the functions $G_\alpha(X, \phi)$
to reproduce nontrivial terms containing the Riemann tensor,
given that the generalized galileon is the most general single-scalar-tensor
theory.\footnote{In
Horndeski's original form of the Lagrangian, the Riemann tensor appears explicitly.}

The most general single-field inflation model (named {\em generalized G-inflation})
was proposed and studied
in Ref.~\cite{Kobayashi:2011nu} based on the Lagrangian~(\ref{GG}).
The purpose of this paper is to explore the multi-field extension
of generalized G-inflation and to discuss to what extent the constructed
multi-field model is general.
We will introduce the multi-galileon theory and its generalization
in the next subsection.

\subsection{Generalized multi-galileons}

The multi-field extension of the galileon
is addressed in Ref.~\cite{Deffayet:2010zh, Bi-galileon}.
Following the same way as the single-field case,
the covariant multi-galileons and their generalization have been found
in Ref.~\cite{Padilla}.
For multiple scalar field $\phi^I$ ($I=1,2,3, ...$), the Lagrangian
is given by
\begin{eqnarray}
{\cal L}&=&G_{2}(X^{IJ}, \phi^K) -G_{3L}(X^{IJ}, \phi^K)\Box\phi^L
+G_4(X^{IJ},\phi^K)R
\nonumber\\&&
+G_{4,\langle IJ\rangle }
\left(\Box\phi^I\Box\phi^J-\nabla_\mu\nabla_\nu\phi^I\nabla^\mu\nabla^\nu\phi^J\right)
+
G_{5 L}(X^{IJ}, \phi^K)G^{\mu\nu}\nabla_\mu\nabla_\nu\phi^L
\nonumber\\&&
-\frac{1}{6} G_{5I, \langle JK\rangle }\left[
\Box\phi^I\Box\phi^J\Box\phi^K
-3\Box\phi^{(I}\nabla_\mu\nabla_\nu\phi^J\nabla^\mu\nabla^\nu\phi^{K)}
+2\nabla_\mu\nabla_\nu\phi^I\nabla^\nu\nabla^\lambda\phi^J\nabla_\lambda\nabla^\mu\phi^K
\right],
\label{Lagrangian}
\end{eqnarray}
where
\begin{eqnarray}
X^{IJ}:=-\frac{1}{2}\partial_\mu\phi^I\partial^\mu\phi^J,
\end{eqnarray}
and we defined the symmetrized derivative
for any functions of $X^{IJ}$ by
\begin{eqnarray}
G_{,\langle IJ\rangle }:=
\frac{1}{2}\left(\frac{\partial G}{\partial X^{IJ}}
+\frac{\partial G}{\partial X^{JI}}\right),
\end{eqnarray}
while hereafter we will use the notation
\begin{eqnarray}
G_{, I}:=\frac{\partial G}{\partial \phi^I}.
\end{eqnarray}
In order for the field equations to be of second order,
one must require that
\begin{eqnarray}
&&
G_{3 IJK}:=G_{3I,\langle JK\rangle},
\quad
G_{4 IJKL}:= G_{4,\langle IJ\rangle,\langle KL\rangle},
\nonumber\\&&
G_{5IJK}:=G_{5I, \langle JK\rangle },\quad
G_{5 IJKLM}:=
G_{5IJK,\langle LM\rangle},
\end{eqnarray}
are symmetric in {\em all} of their indices $I, J, \ldots$.
Hereafter, we will also write $G_{4,\langle IJ\rangle }$ as $G_{4IJ}$,
which is trivially symmetric in $I, J$.
The covariant field equations derived from the Lagrangian~(\ref{Lagrangian})
are presented in the Appendix.
The Lagrangian~(\ref{Lagrangian}) is the starting point and the central focus
of this paper.
We call multi-field inflation based on this Lagrangian
as {\em multi-field G-inflation}.

Padilla and Sivanesan conjectured that
the Lagrangian~(\ref{Lagrangian})
describes the most general multi-scalar-tensor theory
with second-order field equations~\cite{Padilla}.
In contrast to the single-field case, however, at this stage
it is not clear whether this conjecture is true.\footnote{Recently,
Sivanesan proved that in the fixed Minkowski spacetime
and under the condition that the Lagrangian contains up to second
derivatives of $\phi^I$, the most general multiple scalar field theory
with second-order equations of motion is given by the Lagrangian~(\ref{Lagrangian})
with $g_{\mu\nu}\to\eta_{\mu\nu}$~\cite{Sivanesan:2013tba}.}

\section{Background equations}

Let us start with studying the homogeneous and isotropic cosmology
of the generalized multi-galileons. We consider the metric
\begin{eqnarray}
\D s^2=-N^2(t)\D t^2+a^2(t)\delta_{ij}\D x^i\D x^i,
\end{eqnarray}
and homogeneous scalar fields $\phi^I=\phi^I(t)$ (and hence
$X^{IJ}=\dot\phi^I\dot\phi^J/2N^2$, where a dot stands for differentiation with respect to $t$).

Substituting the above ansatz to the action and
varying with respect to $N(t)$, we obtain
the Friedmann equation
\begin{eqnarray}
{\cal E}(\phi^I,\dot\phi^J, H)=0,
\label{Feq}
\end{eqnarray}
where
\begin{eqnarray}
{\cal E}(\phi^I,\dot\phi^J, H)&=&2X^{IJ}G_{2,\langle IJ\rangle}-G_2
+6H\dot\phi^IX^{JK}G_{3IJK}-2X^{IJ}G_{3I,J}
\nonumber\\&&
-6H^2G_4+24H^2X^{IJ}\left(G_{4IJ}+X^{KL}G_{4IJKL}\right)
-12H\dot\phi^IX^{JK}G_{4IJ, K}-6H\dot\phi^IG_{4,I}
\nonumber\\&&
+2H^3\dot\phi^IX^{JK}\left(5G_{5IJK}+2X^{LM}G_{5IJKLM}\right)
-6H^2X^{IJ}\left(
3G_{5I,J}+2X^{KL}G_{5IJK,L}
\right),\label{Friedmann}
\end{eqnarray}
and $H:=\dot a/a$ is the Hubble expansion rate.
(We set $N(t)=1$ after varying the action.)
Now one finds that the basic structure of the (generalized) Friedmann equation~(\ref{Friedmann})
remains unchanged from the single-field counterpart: in addition to the usual $H^2$ term,
the equation can have the terms proportional to $H$ and $H^3$, but no $H^4$ terms or higher powers.

Varying with respect to $a(t)$, we find
\begin{eqnarray}
{\cal P}(\phi^I,\dot\phi^J,\ddot\phi^K, H, \dot H )=0.
\label{dotFriedmann}
\end{eqnarray}
Here, ${\cal P}$ is of the form
\begin{eqnarray}
{\cal P}(\phi^I,\dot\phi^J,\ddot\phi^K, H, \dot H )
&=&\tilde{\cal P}(\phi^I,\dot\phi^J, H)+\ddot\phi^K{\cal B}_K(\phi^I,\dot\phi^J, H)
+2\dot H{\cal G}_{\mathtt T}(\phi^I,\dot\phi^J, H),
\end{eqnarray}
where
\begin{eqnarray}
\tilde {\cal P}&:=&
G_2-2X^{IJ}G_{3I,J}+6H^2G_4
-12H^2X^{IJ}G_{4IJ}
+4H\dot\phi^IG_{4,I}+4X^{IJ}G_{4,I,J}
-8HX^{IJ}\dot\phi^KG_{4IJ,K}
\nonumber\\&&
-4H^3X^{IJ}\dot\phi^K
G_{5IJK}
-4H^2X^{IJ}X^{KL}G_{5IJK,L}+
6H^2X^{IJ}G_{5I,J}
+4HX^{IJ}\dot\phi^K G_{5I,J,K},
\\
{\cal B}_I&:=&
-2X^{JK}G_{3IJK}-4H\dot\phi^JG_{4IJ}-8H X^{JK}\dot\phi^LG_{4IJKL}+2G_{4,I}+4X^{JK}G_{4IJ,K}
\nonumber\\&&
-6H^2X^{JK}G_{5IJK}-4H^2X^{JK}X^{LM}G_{5IJKLM}+4HX^{JK}\dot\phi^LG_{5IJK,L}
+4H\dot\phi^JG_{5(I,J)},
\end{eqnarray}
and
\begin{eqnarray}
{\cal G}_{\mathtt T}:=
2\left[
G_4-2X^{IJ}G_{4 IJ}-X^{IJ}\left(H\dot\phi^K G_{5 IJK}-G_{5I,J}\right)
\right].
\label{def-GT}
\end{eqnarray}
The basic structure is again the same as the single-field counterpart:
second derivatives of the scale factor and the scalar fields appear linearly,
and in particular,
the term ${\cal B}_I$ signals the presence of ``braiding''
between $\phi^I$'s kinetic term and gravity,
which cannot in general be unbraided by a conformal transformation.

The scalar-field equation of motion can be written as
\begin{eqnarray}
\frac{1}{a^3}\frac{\D}{\D t}\left(a^3{\cal J}_I\right)
=\frac{\partial {\cal P}}{\partial \phi^I},
\end{eqnarray}
where
\begin{eqnarray}
{\cal J}_I&:=&\dot\phi^JG_{2,\langle IJ\rangle}+6HX^{JK}G_{3IJK}-2\dot\phi^JG_{3(I,J)}
\nonumber\\&&
-2HG_{4,I}+2\dot\phi^JG_{4,I,J}
+6H^2\dot\phi^JG_{4IJ}+12H^2\dot\phi^JX^{KL}G_{4IJKL}
-8HX^{JK}G_{4JK,I}-12HX^{JK}G_{4IJ,K}
\nonumber\\&&
-6H^2\dot\phi^JG_{5(I,J)}+4HX^{JK}G_{5J,K,I}
+6H^3X^{JK}G_{5IJK}-6H^2G_{5IJK,L}\dot\phi^J X^{KL}-2H^2\dot\phi^JX^{KL}G_{5JKL,I}
\nonumber\\&&
+4H^3X^{JK}X^{KL}G_{5IJKLM}.
\end{eqnarray}
If ${\cal P}$ does not depend on $\phi^I$, then ${\cal J}_I$ is conserved
and the solution is given by ${\cal J}_I = C_I/a^3$,
where $C_I$ is a constant.
Therefore, if ${\cal P}$ does not depend on some of the fields,
the same number of the conserved quantities are present.

Note the relation
\begin{eqnarray}
\dot\phi^I{\cal J}_I={\cal E}+\tilde{\cal P},
\end{eqnarray}
which leads to
\begin{eqnarray}
{\cal E}+{\cal P}=\dot\phi^I{\cal J}_I+\ddot\phi^I{\cal B}_I+2\dot H{\cal G}_{\mathtt T}=0,
\end{eqnarray}
where the last equality follows from Eqs.~(\ref{Feq}) and~(\ref{dotFriedmann}).

\section{Cosmological perturbations}

We now compute the 
quadratic actions for scalar and tensor perturbations.
To do so, we employ the spatially flat gauge and write the perturbed metric as
\begin{eqnarray}
\D s^2=-N^2\D t^2+\gamma_{ij}\left(\D x^i+N^i\D t\right)\left(\D x^j+N^j\D t\right),
\end{eqnarray}
with
\begin{eqnarray}
N=1+\alpha,\quad N_i=\partial_i\beta,\quad \gamma_{ij}=a^2
\left(\delta_{ij}+h_{ij}+\frac{1}{2}h_{ik}h_{kj}\right),
\end{eqnarray}
where $\alpha$ and $\beta$ are scalar perturbations and
$h_{ij}$ is the traceless and transverse tensor perturbation.
The scalar fields are also perturbed as
\begin{eqnarray}
\phi^I= \bar\phi^I(t)+Q^I(t,\mathbf{x}).
\end{eqnarray}
In what follows we will omit the bar on the homogeneous part since there
is no worry about confusion. It should be noted that the tensor
perturbations $h_{ij}$ are gauge-invariant while the scalar
perturbations $\alpha, \beta$, and $Q^{I}$ are gauge-dependent.

\subsection{Tensor perturbations}

The quadratic action for $h_{ij}$ is found to be
\begin{eqnarray}
S_{\mathtt T}^{(2)}
=\frac{1}{8}\int\D t\D^3 x\,a^3\left[
{\cal G}_{\mathtt T}\dot h_{ij}^2-\frac{{\cal F}_{\mathtt T}}{a^2}\partial_kh_{ij}
\partial_kh_{ij}
\right],
\end{eqnarray}
where
\begin{eqnarray}
{\cal F}_{\mathtt T}:=2\left[
G_4-X^{IJ}\left(\ddot\phi^KG_{5IJK}+G_{5I,J}\right)
\right],
\label{def-FT}
\end{eqnarray}
and ${\cal G}_{\mathtt{T}}$ was already defined in Eq.~(\ref{def-GT}).
The propagation speed of the tensor perturbation is given by
$c_{t}^2:={\cal F}_{\mathtt{T}}/{\cal G}_{\mathtt{T}}$.

\subsection{Scalar perturbations}

Expanding the action to second order in scalar perturbations, we obtain
\begin{eqnarray}
S^{(2)}_{\mathtt{S}}&=&\int\D t\D^3x\,a^3\left[\Sigma \alpha^2
-\frac{\partial{\cal E}}{\partial \dot\phi^I}\dot Q^I\alpha
-\frac{\partial{\cal E}}{\partial \phi^I} Q^I\alpha
-\frac{{\cal B}_I}{a^2}\partial^2 Q^I \alpha\right.
+\left({\cal J}_IQ^I
+{\cal B}_I\dot Q^I
-2\Theta\alpha\right)\frac{\partial^2\beta}{a^2}
\nonumber\\
&&\qquad\qquad\quad
\left.+\frac{1}{2}\left({\cal A}_{IJ}\dot Q^I\dot Q^J
+\frac{\partial^2 {\cal P}}{\partial\phi^I \partial \phi^J}Q^IQ^J\right)
+\frac{\partial {\cal J}_J}{\partial\phi^I} Q^I\dot Q^J
-\frac{1}{2a^2}{\cal C}_{IJ}\partial_iQ^I\partial^iQ^J\right],\label{action-original}
\end{eqnarray}
where
\begin{eqnarray}
\Sigma&=&
\frac{1}{2}\left(\dot\phi^I\frac{\partial {\cal E}}{\partial\dot\phi^I}
+H\frac{\partial {\cal E}}{\partial H}\right),
\label{def-Sigma}
\\
\Theta&=&-\frac{1}{6}\frac{\partial{\cal E}}{\partial H},
\label{def-Theta}
\\
{\cal A}_{IJ}&=&
\frac{1}{2}\left(\frac{\partial {\cal J}_I}{\partial\dot\phi^J}
+\frac{\partial {\cal J}_J}{\partial\dot\phi^I}\right)
-\frac{1}{2}\left(\frac{\partial{\cal B}_I}{\partial\phi^J}
+\frac{\partial{\cal B}_J}{\partial\phi^I}\right),
\end{eqnarray}
and
\begin{eqnarray}
{\cal C}_{IJ}&=&
G_{2,\langle IJ\rangle}+\left(\ddot\phi^K+3H\dot\phi^K\right)G_{3IJK}
-2G_{3(I,J)}+a^{-1}\partial_t\left[a \dot\phi^K G_{3IJK}\right]
\nonumber\\
&&+\left(8H^2+6\dot H\right)G_{4IJ}+6H\left(\dot X^{KL}+2HX^{KL}\right)G_{4IJKL}
-4\left(\ddot\phi^K+2H\dot\phi^K\right)G_{4K(I,J)}
\nonumber\\
&&+2a^{-1}\partial_t\left[-aHG_{4IJ}+4aHX^{KL}G_{4IJKL}-a\dot\phi^KG_{4IJ,K}\right]
\nonumber\\
&&
-2\left(3H^2+2\dot H\right)G_{5(I,J)}+\left(
3H^2\ddot\phi^K+5H^3\dot\phi^K+6H\dot H\dot\phi^K\right)G_{5IJK}
\nonumber\\
&&-4H\left(\dot X^{KL}+HX^{KL}\right)G_{5KL(I,J)}
+\left(3H^2\dot X^{KL}\dot\phi^M+2H^3X^{KL}\dot\phi^M\right)G_{5IJKLM}
\nonumber\\
&&
+a^{-1}\partial_t\left[
-aH^2\dot\phi^KG_{5IJK}-4aHX^{KL}G_{5IJK,L}
+2aH^2X^{KL}\dot\phi^MG_{5IJKLM}
\right].\label{CIJequation}
\end{eqnarray}
It is worth emphasizing that
all the coefficients other than ${\cal C}_{IJ}$
can be computed directly from the quantities appearing in the background equations of motion.
If, for example, the Lagrangian contains $(\delta_{IJ}X^{IJ})^2-\delta_{IJ}\delta_{KL}X^{IK}X^{JL}$,
its effect can only be probed via ${\cal C}_{IJ}$
since $(\delta_{IJ}X^{IJ})^2-\delta_{IJ}\delta_{KL}X^{IK}X^{JL}=0$
in a homogeneous background.

It can be seen that the following identities hold:
\begin{eqnarray}
2X^{IJ}{\cal C}_{IJ}&=&{\cal E}+{\cal P}
+2\left[
2H\left(\dot{\cal G}_{\mathtt T}+H{\cal G}_{\mathtt T}\right)
-\left(\dot\Theta+H\Theta\right)-H^2{\cal F}_{\mathtt T}
\right],\label{Id1}
\\
\frac{\partial{\cal E}}{\partial \dot\phi^I}
&=&\dot\phi^J{\cal A}_{IJ}-3H{\cal B}_I,\label{Id2}
\\
2\Theta &=& \dot\phi^I{\cal B}_I+2H{\cal G}_{\mathtt T}.\label{Id3}
\end{eqnarray}
Note that Eqs.~(\ref{Id2}) and~(\ref{Id3}) are combined to give
\begin{eqnarray}
X^{IJ}{\cal A}_{IJ}=\Sigma +6H\Theta-3H^2{\cal G}_{\mathtt{T}}.
\end{eqnarray}

Varying the action with respect to $\alpha$ and $\beta$,
we obtain the constraint equations,
\begin{eqnarray}
2\Sigma \alpha
-\frac{\partial{\cal E}}{\partial\dot\phi^I}\dot Q^I
-\frac{\partial{\cal E}}{\partial\phi^I}Q^I
-{\cal B}_I\frac{\partial^2Q^I}{a^2}-2\Theta\frac{\partial^2 \beta}{a^2}
&=&0,\label{energy-const}
\\
{\cal J}_IQ^I+{\cal B}_I\dot Q^I-2\Theta\alpha&=&0.\label{momentum-const}
\end{eqnarray}
Substituting Eq.~(\ref{momentum-const}) to the action,
we arrive at
\begin{eqnarray}
S_{\mathtt{S}}^{(2)}=\frac{1}{2}\int\D t\D^3x\,a^3\left[
{\cal K}_{IJ}\dot Q^I\dot Q^J-\frac{1}{a^2}{\cal D}_{IJ}\partial_iQ^I\partial^iQ^J
-{\cal M}_{IJ}Q^IQ^J+2\Omega_{IJ} Q^I \dot Q^J
\right],\label{final-action}
\end{eqnarray}
where
\begin{eqnarray}
{\cal K}_{IJ}&=&{\cal A}_{IJ}+
\frac{{\cal B}_I{\cal B}_J}{2\Theta^2}\left(\Sigma +6H\Theta\right)
-\frac{\dot\phi^K}{\Theta}{\cal B}_{(I}{\cal A}_{J)K},
\\
{\cal D}_{IJ}&=&{\cal C}_{IJ}-\frac{1}{\Theta}{\cal J}_{(I}{\cal B}_{J)}
+\frac{1}{a}\frac{\D}{\D t}\left(\frac{a{\cal B}_I{\cal B}_J}{2\Theta}\right),
\\
{\cal M}_{IJ}&=&-\frac{\partial^2{\cal P}}{\partial\phi^I\partial\phi^J}
-\frac{\Sigma}{2\Theta^2}{\cal J}_I{\cal J}_J
+\frac{1}{\Theta}\frac{\partial{\cal E}}{\partial\phi^{(I}}
{\cal J}_{J)},
\end{eqnarray}
and
\begin{eqnarray}
\Omega_{IJ}=\frac{\partial{\cal J}_J}{\partial\phi^I}
-\frac{1}{2\Theta}\frac{\partial{\cal E}}{\partial\phi^I}{\cal B}_J
+\frac{{\cal J}_I{\cal B}_J}{2\Theta^2}\left(
\Sigma+3H\Theta
\right)-\frac{\dot\phi^K}{2\Theta}{\cal J}_{I}{\cal A}_{JK}.
\end{eqnarray}
In deriving the above expression we have used Eq.~(\ref{Id2})
to remove $\partial{\cal E}/\partial\dot\phi^I$.
The equation of motion derived from the action~(\ref{final-action}) is given by
\begin{eqnarray}
{\cal K}_{IJ}\ddot Q^J-\frac{1}{a^2}{\cal D}_{IJ}\partial^2Q^J
+\left(\dot{\cal K}_{IJ}+3H{\cal K}_{IJ}+\Omega_{JI}-\Omega_{IJ}\right)\dot Q^J
+\left({\cal M}_{IJ}+\dot\Omega_{JI}+3H\Omega_{JI}\right)Q^J=0.
\end{eqnarray}
To avoid ghost and gradient instabilities, we impose that
all the eigenvalues of the matrices ${\cal K}_{IJ}$ and ${\cal D}_{IJ}$
are positive.

Before closing this section, let us give a short remark
on the relation between the above result obtained 
in the spatially flat gauge and the single-field result ($\phi^{1}=\phi$) 
derived previously in the unitary gauge~\cite{Kobayashi:2011nu}.
In the unitary gauge, $\phi = \bar\phi(t)$, that is, $Q(t,\mathbf{x})=0$,
and the perturbed metric is written as
\begin{eqnarray}
\D s^2=-\widetilde{N}^2\D t^2+\widetilde{\gamma}_{ij}
\left(\D x^i+\widetilde{N}^i\D t\right)\left(\D x^j+\widetilde{N}^j\D t\right),
\end{eqnarray}
with
\begin{eqnarray}
\widetilde{N}=1+\widetilde{\alpha},\quad
 \widetilde{N}_i=\partial_i\widetilde{\beta},\quad \gamma_{ij}=a^2 e^{-2{\cal R}}
\left(\delta_{ij}+h_{ij}+\frac{1}{2}h_{ik}h_{kj}\right),
\end{eqnarray}
where $\widetilde{\alpha}$, $\widetilde{\beta}$, and ${\cal R}$ are scalar
perturbations, and
$h_{ij}$ is the traceless and transverse tensor
perturbation. The tensor perturbation is invariant
under the gauge transformation.
By the gauge transformation from the spatially flat gauge (${\cal R}=0$) 
to the unitary gauge ($Q=0$), the scalar perturbations transform as
\begin{eqnarray}
  \alpha&\to&\widetilde{\alpha} = \alpha - \frac{\D}{\D t}
   \left(\frac{Q}{\dot{\phi}}\right), \\
\beta&\to&  \widetilde{\beta} = \beta +
   \frac{Q}{\dot{\phi}}, \\
  0&\to&{\cal R} =0+ H \frac{Q}{\dot{\phi}}.
\end{eqnarray}
%
Inserting
these into Eq.~(\ref{action-original}) for the single-field
case, we can easily find the second-order action for the scalar
perturbations in the unitary gauge,
%
\begin{equation}
S^{(2)}_{\mathtt S} =\int \D t\D^3 x a^3\biggl[
-3{\cal G}_{\mathtt T}\dot{\cal R}^2+\frac{{\cal F}_{\mathtt T}}{a^2}
\partial_i {\cal R} \partial^i {\cal R}
+\Sigma \widetilde{\alpha}^2
-2\Theta\widetilde{\alpha}\frac{\partial^2 \widetilde{\beta}}{a^2}
-2{\cal G}_{\mathtt T}\dot {\cal R}\frac{\partial^2\widetilde{\beta}}{a^2}
-6\Theta \widetilde{\alpha}\dot{\cal R}+2{\cal G}_{\mathtt T}
\widetilde{\alpha}\frac{\partial^2{\cal R}}{a^2}
\biggr],\label{scalarunitary}
\end{equation}
where ${\cal F}_{\mathtt T}, {\cal G}_{\mathtt T}, \Sigma$, and $\Theta$ are the single-field
counterparts of Eqs.~(\ref{def-GT}), (\ref{def-FT}),
(\ref{def-Sigma}), and (\ref{def-Theta}). This expression completely
coincides with that obtained in Ref.~\cite{Kobayashi:2011nu} after changing
the notation as ${\cal R}\to -\zeta$.

\subsection{Two-field model}

For simplicity, let us now focus on a two-field model of inflation, $I=1, 2$.
We introduce the field space metric $\mathfrak{g}_{IJ}(\phi^K)$,\footnote{It
is worth stressing that the formulation so far does not require
utilizing a field space metric; here we introduce it for the first time.}
by which
we measure the size of instantaneous adiabatic and entropy modes.
Those modes are defined, respectively, as the perturbations parallel and perpendicular
to the background trajectory in field space, and thus it is convenient to
employ the basis vectors $e_\sigma^I$ and $e_s^I$ defined by
$e^I_\sigma :=\dot\phi^I/\dot \sigma:=\dot\phi^I/\sqrt{2\mathfrak{g}_{JK}X^{JK}}$,
$\mathfrak{g}_{IJ}e_s^Ie_s^J=1$, and
$\mathfrak{g}_{IJ}e_\sigma^Ie_s^J=0$~\cite{Gordon:2000hv}.
In terms of $\{ e_\sigma^I, e_s^I\}$, the perturbation $Q^I$
can be decomposed into the adiabatic mode, $Q^I_\sigma$,
and the entropy mode, $Q^I_s$, as
\begin{eqnarray}
Q^I=Q_\sigma e_\sigma^I+Q_s e^I_s.
\end{eqnarray}
We then define the curvature and the normalized entropy perturbations, respectively, as
\begin{eqnarray}
{\cal R}:=\frac{H}{\dot\sigma} Q_\sigma,
\quad
{\cal S}:=\frac{H}{\dot\sigma} Q_s,
\end{eqnarray}
in terms of which the quadratic action reduces to
\begin{eqnarray}
S=\int\D t\D^3x\,a^3\left[
{\cal G}_{\mathtt S}\dot{{\cal R}}^2-\frac{{\cal F}_{\mathtt{S}}}{a^2}(\partial{\cal R})^2
+{\cal L}_{{\cal SS}}+{\cal L}_{{\cal R}{\cal S}}
\right].
\end{eqnarray}
Here,
\begin{eqnarray}
{\cal G}_{\mathtt{S}}&:=&
\frac{X}{H^2}
{\cal K}_{IJ}e_\sigma^Ie_\sigma^J
=
\frac{\Sigma }{\Theta^2}{\cal G}_{\mathtt{T}}^2+3{\cal G}_{\mathtt{T}},
\\
{\cal F}_{\mathtt{S}}&:=&
\frac{X}{H^2}
{\cal D}_{IJ}e_\sigma^Ie_\sigma^J
=
\frac{1}{a}\frac{\D}{\D t}\left(\frac{a}{\Theta}{\cal G}_{\mathtt{T}}^2\right)
-{\cal F}_{\mathtt{T}},
\end{eqnarray}
${\cal L}_{{\cal SS}}\supset \dot {\cal S}^2,\; (\partial{\cal S})^2,\;{\cal S}^2$,
and ${\cal L}_{{\cal RS}}\supset \dot {\cal R}\dot{\cal S},\;\partial{\cal R}\partial{\cal S},\;\ldots$.
Explicit expressions for ${\cal L}_{\cal SS}$ and ${\cal L}_{\cal RS}$
are not illuminating.
Turning off the entropy mode, one can verify that
the above action coincides with the single-field result~\cite{Kobayashi:2011nu}.
Of course, in general, requiring ${\cal G}_{\mathtt S}>0$ and ${\cal F}_{\mathtt S}>0$ is
not sufficient to avoid instabilities.

On superhorizon scales where spatial gradients may be neglected, one can show using
the Hamiltonian and momentum constraints~(\ref{energy-const}),~(\ref{momentum-const}) that
\begin{eqnarray}
\frac{\Theta{\cal G}_{\mathtt S}}{{\cal G}_{\mathtt T}}\dot{{\cal R}}\simeq
I(Q_s, \dot Q_s),\label{dotR}
\end{eqnarray}
where
\begin{eqnarray}
I(Q_s, \dot Q_s):=
\frac{\Sigma}{2\Theta}\left[
{\cal B}_I\partial_t\left(Q_se^I_s\right)+{\cal J}_IQ_se^I_s
\right]
-\frac{1}{2}\frac{\partial{\cal E}}{\partial\dot\phi^I}\partial_t\left(Q_se^I_s\right)
-\frac{1}{2}\frac{\partial{\cal E}}{\partial\phi^I}Q_se^I_s.
\end{eqnarray}
This implies the generic conclusion in multi-field models that
the superhorizon curvature perturbation does not stay constant
in the presence of the entropy perturbations,
as first demonstrated in Ref.~\cite{Starobinsky:1994mh}.
The coupled superhorizon evolution equations for ${\cal R}$ and ${\cal S}$
can be written in the form
\begin{eqnarray}
&&{\cal A}_1\frac{\D^2}{\D t^2}\left( \begin{array}{cc} {\cal R}\\ {\cal S}\\ \end{array} \right)
+{\cal A}_2\frac{\D}{\D t}\left( \begin{array}{cc} {\cal R}\\ {\cal S}\\ \end{array} \right)
+{\cal A}_3\left( \begin{array}{cc} {\cal R}\\ {\cal S}\\ \end{array} \right)\simeq 0
\nonumber\\
&\Rightarrow&
{\cal A}_3^{-1}
{\cal A}_1\frac{\D^2}{\D t^2}\left( \begin{array}{cc} {\cal R}\\ {\cal S}\\ \end{array} \right)
+{\cal A}_3^{-1}
{\cal A}_2\frac{\D}{\D t}\left( \begin{array}{cc} {\cal R}\\ {\cal S}\\ \end{array} \right)
+\left( \begin{array}{cc} {\cal R}\\ {\cal S}\\ \end{array} \right)\simeq 0,
\end{eqnarray}
where ${\cal A}_1, {\cal A}_2,$ and ${\cal A}_3$ are some $2\times 2$ matrices.
Substituting Eq.~(\ref{dotR}) to the second equation, one sees that
the superhorizon evolution of the entropy perturbation is
independent of ${\cal R}$. This fact holds quite generically.
On the other hand, it is difficult to derive a generic conclusion about
the evolution of the adiabatic and entropy modes
on subhorizon scales, as the two modes are coupled in a model-dependent way.

\section{Is multi-field DBI galileon inflation included in multi-field G-inflation?}

\subsection{Galileons from an embedded brane}

In Ref.~\cite{deRham:2010eu}, the galileon field is reformulated
as a position modulus of
a probe brane embedded in a five-dimensional bulk (see also Ref.~\cite{Goon:2011uw}).
The derivation is similar to that used to obtain the DBI scalar field.
Suppose that the Lagrangian for the probe brane contains
an induced Einstein-Hilbert term: ${\cal L}_{\rm brane}\supset \sqrt{-\gamma}R[\gamma]$.
Substituting to ${\cal L}_{\rm brane}$
the induced metric $\gamma_{\mu\nu}=g_{\mu\nu}+f\partial_\mu\phi\partial_\nu\phi$,
where $f$ is assumed to be constant here for simplicity,
one obtains the following term up to a total derivative:
\begin{eqnarray}
\sqrt{-\gamma}R[\gamma]=\sqrt{-g}\left\{\sqrt{1-2f X}R[g]
-\frac{f}{\sqrt{1-2fX}}\left[(\Box\phi)^2-(\nabla_\mu\nabla_\nu\phi)^2\right]\right\}.
\end{eqnarray}
From this it is clear that the {\em single} DBI galileon
is a subclass of the generalized galileon~(\ref{GG})
corresponding to $G_4=\sqrt{1-2fX}$.


A multi-field generalization of the DBI galileon has been demonstrated
in Ref.~\cite{Hinterbichler:2010xn} using a higher codimension bulk,
and cosmology of the multi-field DBI galileons has been addressed
in Refs.~\cite{RenauxPetel:2011dv, RenauxPetel:2011uk, Koyama:2013wma}.
Following Refs.~\cite{RenauxPetel:2011dv, RenauxPetel:2011uk},
let us consider the action
\begin{eqnarray}
S=\int\D^4x \sqrt{-\gamma}\left[
-\frac{1}{f}+\frac{M^2}{2}R[\gamma]
\right],\label{4deff}
\end{eqnarray}
where $M$ is some mass scale, and
substitute the induced metric
\begin{eqnarray}
\gamma_{\mu\nu}=g_{\mu\nu}+f\delta_{IJ}\partial_\mu\phi^I\partial_\nu\phi^J,
\end{eqnarray}
to the above action.
We assume the flat field space metric, ${\mathfrak g}_{IJ}=\delta_{IJ}$,
and a constant warp factor, $f=\,$const.
The first term in the action~(\ref{4deff})
leads to a specific case of $G_2(X^{IJ}, \phi^K)$, 
and its property
has been studied extensively in the context of usual multi-field DBI
inflation~\cite{Langlois:2008wt, Langlois:2008qf}.
Our main interest is thus the second term in Eq.~(\ref{4deff}).
Note in passing that in Refs.~\cite{RenauxPetel:2011dv, RenauxPetel:2011uk}
the Einstein-Hilbert term for the cosmological metric, $\sqrt{-g}R[g]$,
is included in the total action in addition to the induced Einstein-Hilbert
term $\sqrt{-\gamma}R[\gamma]$. Since the term $\sqrt{-g}R[g]$
just adds a constant contribution to $G_4(X^{IJ}, \phi^K)$,
its role is trivial formally. For this reason, we omit
the standard Einstein-Hilbert term from the action~(\ref{4deff}).


A detailed computation of the explicit multi-field action from $\sqrt{-\gamma}R[\gamma]$
is presented in Appendix A of Ref.~\cite{RenauxPetel:2011uk}.
The resultant action is quite complicated, which hinders
comparing the multi-field DBI galileons with the Lagrangian~(\ref{Lagrangian})
to determine (if possible)
the corresponding functions $G_{3I}, G_4$, and $G_{5I}$.
To simplify the analysis, we expand
the action to second order in $f\partial_\mu\phi^I\partial_\nu\phi^J$.
We then obtain, up to a total derivative,
\begin{eqnarray}
\sqrt{-\gamma}R[\gamma]=\sqrt{-g}\left[
R[g]+f{\cal L}^{(1)}+f^2{\cal L}^{(2)}+{\cal O}(f^3)\right],
\end{eqnarray}
where
\begin{eqnarray}
{\cal L}^{(1)}&=&
-XR[g]-\nabla^\mu\phi_I\nabla^\nu\phi^IR_{\mu\nu}[g],
\\
{\cal L}^{(2)}&=&\frac{1}{2}\left(X^2-2X_{IJ}X^{IJ}\right)R[g]
+\left(X\delta_{IJ}-2X_{IJ}\right)\nabla^\mu\phi^I\nabla^\nu\phi^JR_{\mu\nu}[g]
\nonumber\\&&
+\partial_\mu\phi^I\partial_\nu\phi^J\left(
\nabla^\nu\nabla_\lambda\phi_I\nabla^\mu\nabla^\lambda \phi_J-\nabla^\mu\nabla^\nu\phi_I\Box\phi_J
\right),
\end{eqnarray}
and we write $\nabla_\mu\phi_I=\delta_{IJ}\nabla_\mu\phi^J$,
$X_{IJ}=\delta_{IK}\delta_{JL}X^{KL}$, and $X=\delta_{IJ}X^{IJ}$.
A further manipulation shows that, up to a total derivative,
\begin{eqnarray}
{\cal L}^{(1)}=-XR[g]-\delta_{IJ}\left(\Box\phi^I\Box\phi^J-\nabla_\mu\nabla_\nu\phi^I
\nabla^\mu\nabla^\nu\phi^J\right),
\end{eqnarray}
and
\begin{eqnarray}
{\cal L}^{(2)}&=&-\frac{1}{6}\left(X^2+2X_{IJ}X^{IJ}\right)R
-\frac{1}{3}\left(X\delta_{IJ}+2X_{IJ}\right)
\left(\Box\phi^I\Box\phi^J-\nabla_\mu\nabla_\nu\phi^I
\nabla^\mu\nabla^\nu\phi^J\right)
\nonumber\\&&
+\frac{1}{3}L^{\mu\alpha\nu\beta}\partial_\mu\phi_I
\partial_\nu\phi^I\partial_\alpha\phi_J\partial_\beta\phi^J,
\end{eqnarray}
where $L^{\mu\alpha\nu\beta}$ is the double dual Riemann tensor
defined in Eq.~(\ref{def-ddriem}).
Thus, we see that
$\sqrt{-\gamma}R[\gamma]$ can be written in terms of
the generalized multi-galileons with
\begin{eqnarray}
G_4=1-f\delta_{IJ}X^{IJ}-\frac{f^2}{6}\left(
\delta_{IJ}\delta_{KL}
+\delta_{IK}\delta_{JL}+\delta_{IL}\delta_{JK}
\right)X^{IJ}X^{KL}+{\cal O}(f^3),\label{G4f2}
\end{eqnarray}
plus
\begin{eqnarray}
{\cal L}_{\ast }:=\frac{f^2}{3}\delta_{IJ}\delta_{KL}L^{\mu\alpha\nu\beta}\partial_\mu\phi^I
\partial_\nu\phi^J\partial_\alpha\phi^K\partial_\beta\phi^L+{\cal O}(f^3).
\end{eqnarray}
It is obvious from Eq.~(\ref{G4f2}) that $G_{4IJKL}$ is symmetric in its indices $I, J, K, L$.

Notice that the extra term ${\cal L}_\ast$ is {\em not} the
multi-field version of (\ref{DDRiem}). This term
only manifests itself provided that spacetime is curved
{\em and} there are multiple scalar fields.
Apparently, ${\cal L}_\ast$ is not of the form of the generalized multi-galileons.
However, one must be careful to conclude that
this term is not included in the Lagrangian of the generalized multi-galileons,
because in the single-field case terms such as~(\ref{NMGB}) and~(\ref{DDRiem}) can be
recast in the form of the generalized galileon in a nontrivial manner.
Using the concrete example of the cosmological setup shown above,
we explore whether or not
${\cal L}_*$ can be reproduced by choosing nontrivial $G_4$ and other functions 
in the next section.

\subsection{Multi-field DBI galileons versus generalized multi-galileons}

In a cosmological setting, ${\cal L}_\ast$ has a distinctive feature.
Since $L^{0000}=0$ due to its antisymmetric nature,
this term does not contribute to the background equations.
We also have $L^{000i}=0$, so that ${\cal L}_\ast$ gives rise to only
$\partial_iQ^I\partial^iQ^J$ in the quadratic action.
Therefore, for our purpose
it is sufficient to evaluate $L^{0i0j}$ at zeroth order:
$L^{0i0j}=-L^{0ij0}=-(H^2/a^2)\delta^{ij}$. Thus,
up to quadratic order in cosmological perturbations,
${\cal L}_\ast$ simply reads
\begin{eqnarray}
{\cal L}_\ast=\frac{4}{3}f^2H^2\left(X_{IJ}-\delta_{IJ}X\right)\partial_iQ^I\partial^iQ^J
+{\cal O}(f^3).
\end{eqnarray}
This then gives the additional contribution to the action~(\ref{action-original}):
${\cal C}_{IJ}\to {\cal C}_{IJ}+\Delta{\cal C}_{IJ}$, where
\begin{eqnarray}
\Delta {\cal C}_{IJ}= -\frac{4M^2}{3}f^2H^2\left(X_{IJ}-\delta_{IJ}X\right),\label{CLast}
\end{eqnarray}
without modifying any other coefficients
and the background equations
at ${\cal O}(f^2)$.
The relevant terms to be compared with Eq.~(\ref{CLast})
are those proportional to $H^2$ in the formula~(\ref{CIJequation}),
\begin{eqnarray}
{\cal C}_{IJ}&\supset&
6H^2G_{4IJ}+20H^2X^{KL}G_{4IJKL}
\nonumber\\&&
-6H^2G_{5(I,J)}-4H^2X^{KL}G_{5KL(I,J)}-6H^2X^{KL}G_{5IJK,L}+4H^2X^{KL}X^{MN}G_{5IJKLM,N}.
\end{eqnarray}
We see that $G_2$ and $G_3$ are irrelevant to $\Delta {\cal C}_{IJ}$.
To reproduce Eq.~(\ref{CLast}) by $G_4$ and $G_{5I}$, additional contributions in
$G_{4IJ}$ and $G_{5(I,J)}$ must be linear in $X^{IJ}$.
Since $G_{4IJKL}$ and $G_{5IJ(K,L)}$ must be symmetric in $I, J, K, L$,
the only possible choice is written as
$G_{4IJ}\supset g_4\left(\delta_{IJ}\delta_{KL}+\delta_{IK}\delta_{JL}+\delta_{IL}\delta_{JK}\right)X^{KL}$
and
$G_{5(I,J)}\supset g_5\left(\delta_{IJ}\delta_{KL}+\delta_{IK}\delta_{JL}+\delta_{IL}\delta_{JK}\right)X^{KL}$,
where $g_4$ and $g_5$ are some constants.
However, these two terms can never be combined to give Eq.~(\ref{CLast}).
We therefore conclude that the extra term ${\cal L}_\ast$
cannot be described by any consistent choice of $G_2, G_{3I}, G_4, G_{5I}$,
and hence the generalized multi-galileon theory is {\em not} the
most general multi-scalar-tensor theory with second-order field equations.

Note that, since $\Delta{\cal C}_{IJ}e_\sigma^Ie_\sigma^J=0$,
this additional term has no impact on the instantaneous adiabatic mode.
By a straightforward calculation, one can confirm that
our formulas with Eq.~(\ref{G4f2}) reproduce
the result of Ref.~\cite{RenauxPetel:2011uk} (up to ${\cal O}(f^2)$)
except only for the coefficient of $\partial_iQ^s\partial^iQ^s$.
Of course, this coefficient can also be correctly reproduced
by taking $\Delta {\cal C}_{IJ}$ into account.

\section{Conclusions}

In this paper, we have formulated cosmological perturbation theory in
inflation models with generalized multi-galileons.  The generalized
multi-galileon theory is constructed in such a way that the
multi-galileons in the fixed flat spacetime are covariantized while
maintaining second-order field equations.  The resultant inflation model
is more general than the multi-field models considered in the
literature, including multi-DBI inflation and multi-field inflation with
nonminimal couplings. Multi-field G-inflation allows us to treat
those different models on equal footing, and therefore is useful for
testing multi-field inflation models against cosmological observations.
The generalized G-inflation approach is in contrast to, and
complementary to, the effective field theory approach to
inflation~\cite{eftinf,multieft}, as the guiding principle of the former
is the second-order field equations free of Ostrogradski's ghost, rather
than symmetry.\footnote{The detailed relation between the
generalized Galileon and the effective field theory approach is
discussed, for example, in Ref. \cite{Gleyzes:2013ooa}.} In this
generalized setup we have demonstrated the fact that the superhorizon
evolution of the curvature perturbation is affected by the entropy
perturbations, while the entropy perturbations evolve independently of
the curvature perturbation on superhorizon scales.  It is, however,
beyond the scope of the present paper to address generically the
generation and evolution of those perturbations on subhorizon scales
because the two modes are coupled in a model-dependent way.  This point
is left for future study.

We have also inspected the question of whether or not the generalized
multi-galileon theory is the most general multi-scalar-tensor theory,
{\em i.e.}, the multi-field version of the Horndeski theory.
Unfortunately, the answer is {\em no}.  To present a counterexample, we
have considered the so-called multi-field DBI galileons derived from an
induced gravity term of an embedded probe brane.  Comparing cosmological
perturbation equations, we have shown that the multi-DBI galileons give
rise to an extra term that cannot be reproduced by any consistent choice of
arbitrary functions in the generalized multi-galileon theory.
Therefore, it would be interesting and desirable to explore the truly most
general multi-scalar-tensor theory in a systematic way.

\acknowledgments
We thank Xian Gao for helpful comments.
This work was supported in part by JSPS Grant-in-Aid for Young
Scientists (B) No.~24740161 (T.K.), the Grant-in-Aid for Scientific
Research on Innovative Areas No.~24111706 (M.Y.), and the Grant-in-Aid
for Scientific Research No.~25287054 (M.Y.).  
The work of N.T. is supported in part the by World Premier International 
Research Center Initiative (WPI Initiative), MEXT, Japan, 
and JSPS Grant-in-Aid for Scientific Research 25$\cdot$755. 

\appendix
\section{Covariant equations of motion}

For completeness we derive the covariant equations of motion
for the generalized multi-galileons.
Variation of the action with respect to $g_{\mu\nu}$ and $\phi^I$ leads to
\begin{eqnarray}
\delta\left({\sqrt{-g}{\cal L}}\right)
=\frac{1}{2}\sqrt{-g}\,{\cal G}_{\mu\nu}\delta g^{\mu\nu}
+\sqrt{-g}\left[P_I
-\nabla_\mu  J_I^\mu\right]\delta\phi^I,
\end{eqnarray}
where
\begin{eqnarray}
{\cal G}_{\mu\nu}&=&-G_{2,\langle IJ\rangle}\nabla_\mu\phi^I\nabla_\nu\phi^J-g_{\mu\nu}G_2
+G_{3IJK}\Box\phi^I\nabla_\mu\phi^I\nabla_\nu\phi^J
+2\nabla_{(\mu}G_{3I}\nabla_{\nu)}\phi^I-g_{\mu\nu}\nabla^\lambda G_{3I}\nabla_\lambda\phi^I
\nonumber\\&&
+2G_4G_{\mu\nu}-G_{4IJ}R\nabla_{\mu}\phi^I\nabla_{\nu }\phi^J
+2G_{4IJ}R_{\mu\alpha\nu\beta}\nabla^\alpha\phi^I\nabla^\beta\phi^J+
4G_{4IJ}R_{\mu\lambda}\nabla_\nu\phi^I\nabla^\lambda\phi^J
\nonumber\\&&
+g_{\mu\nu} G_{4IJ}\left(\Box\phi^I\Box\phi^J-\nabla_\alpha
\nabla_\beta\phi^I\nabla^\alpha\nabla^\beta\phi^J\right)
+2g_{\mu\nu}\left(G_{4,I}\Box\phi^I+\nabla_\lambda G_{4,I}\nabla^\lambda\phi^I
+\nabla_\lambda G_{4IJ}\nabla^\lambda\phi^I\Box\phi^J\right)
\nonumber\\&&
-2\left(G_{4,I}\nabla_\mu\nabla_\nu\phi^I+\nabla_\mu G_{4,I}\nabla_\nu\phi^I
+G_{4IJ}\Box\phi^I\nabla_\mu\nabla_\nu\phi^J
-G_{4IJ}\nabla_\mu\nabla_\lambda\phi^I\nabla_\nu\nabla^\lambda\phi^J\right)
-2\nabla_\lambda G_{4IJ}\nabla^\lambda\phi^I\nabla_\mu\nabla_\nu\phi^J
\nonumber\\&&
-4\nabla_\mu G_{4IJ}\nabla_\nu\phi^I\Box\phi^J
+4\nabla_\lambda G_{4IJ}\nabla_\mu\phi^I\nabla_\nu\nabla^\lambda\phi^J
-G_{4IJKL}\nabla_\mu\phi^K\nabla_\nu\phi^L\left(
\Box\phi^I\Box\phi^J-\nabla_\alpha
\nabla_\beta\phi^I\nabla^\alpha\nabla^\beta\phi^J
\right)
\nonumber\\&&
-2g_{\mu\nu}\nabla_\alpha G_{4IJ}\nabla^\alpha\nabla^\beta\phi^I\nabla_\beta\phi^J
+2\nabla_\mu G_{4IJ}\nabla_\nu\nabla^\lambda\phi^I\nabla_\lambda\phi^J
-2g_{\mu\nu}
G_{4IJ}R_{\alpha\beta}\nabla^\alpha\phi^I\nabla^\beta\phi^J
\nonumber\\&&
+2G_{5IJK}\biggl[R^{\alpha\beta}\nabla_\alpha\phi^I\nabla_\mu\phi^J\nabla_\nu\nabla_\beta\phi^K
-\frac{1}{2}R^{\alpha\beta}\nabla_\alpha\phi^I\nabla_\beta\phi^J\nabla_\mu\nabla_\nu\phi^K
-R_{\alpha\mu}\nabla_\nu\phi^I\nabla^\alpha\phi^J\Box\phi^K
\nonumber\\&&
-\frac{1}{2}R_{\alpha\mu\beta\nu}\nabla^\alpha\phi^I\nabla^\beta\phi^J\Box\phi^K
+R_{\alpha\mu\beta\lambda}\nabla_\nu\phi^I\nabla^\lambda\phi^J\nabla^\alpha\nabla^\beta\phi^K
+R_{\alpha\mu\beta\lambda}\nabla^\alpha\phi^I\nabla^\beta\phi^J\nabla_\nu\nabla^\lambda\phi^K\biggr]
\nonumber\\&&
+2G_{5(I,J)}\left(\Box\phi^I\nabla_\mu\nabla_\nu\phi^J-\nabla_\mu\nabla_\lambda\phi^I
\nabla_\nu\nabla^\lambda\phi^J\right)
+\nabla_\lambda G_{5I,J}\nabla_\mu\nabla_\nu\phi^I\nabla^\lambda\phi^J
+\nabla_\mu G_{5I,J}\Box\phi^I\nabla_\nu\phi^J
\nonumber\\&&
-2\nabla_\mu G_{5I,J}\nabla_\nu\nabla_\lambda\phi^I
\nabla^\lambda\phi^J
-2\nabla^\alpha G_{5I}\nabla^\beta\phi^IR_{\mu\alpha\nu\beta}
+\nabla_\lambda G_{5I}\nabla^\lambda\phi^I G_{\mu\nu}
-2\nabla_\mu G_I\nabla^\lambda\phi^IR_{\nu\lambda}
\nonumber\\&&
-2\nabla_\lambda G_{5I}\nabla_\mu\phi^I G_{\nu\lambda}
-G_{5IJK}\nabla_\mu\phi^I\nabla_\nu\phi^J\nabla_{\alpha}\nabla_\beta\phi^K G^{\alpha\beta}
+G_{5IJK}\left(\Box\phi^I\Box\phi^J-\nabla_\alpha\nabla_\beta\phi^I\nabla^\alpha\nabla^\beta\phi^J
\right)\nabla_\mu\nabla_\nu\phi^K
\nonumber\\&&
-2G_{5IJK}\left(\Box\phi^I\nabla^\lambda\nabla_\mu\phi^J\nabla_\lambda\nabla_\nu\phi^K
-\nabla^\alpha\nabla^\beta\phi^I\nabla_\alpha\nabla_\mu\phi^J\nabla_\beta\nabla_\nu\phi^K\right)
-\nabla_\alpha G_{5IJK}\nabla_\beta\phi^I\nabla^\alpha\nabla^\beta\phi^J\nabla_\mu\nabla_\nu\phi^K
\nonumber\\&&
+\nabla_\alpha G_{5IJK}\nabla^\alpha\phi^I\left(
\Box\phi^J\nabla_\mu\nabla_\nu\phi^K-\nabla_\beta\nabla_\mu\phi^J\nabla^\beta\nabla_\nu\phi^K
\right)
+2\nabla_\mu G_{5IJK}\nabla_\nu\nabla_\alpha\phi^I\nabla^\alpha\nabla^\beta\phi^J\nabla_\beta\phi^K
\nonumber\\&&
-2\nabla_\alpha G_{5IJK}
\left(
\Box\phi^I\nabla^\alpha\nabla_\mu\phi^J-\nabla^\alpha\nabla^\beta\phi^I
\nabla_\beta\nabla_\mu\phi^J\right)\nabla_\nu\phi^K
+
\nabla_\mu G_{5IJK}\left(\Box\phi^I\Box\phi^J-\nabla_\alpha\nabla_\beta\phi^I
\nabla^\alpha\nabla^\beta\phi^J\right)\nabla_\nu\phi^K
\nonumber\\&&
-\nabla_\mu G_{5IJK}\nabla_\nu\nabla_\lambda\phi^I\Box\phi^J\nabla^\lambda\phi^K
+\frac{1}{6}G_{5IJKLM}\bigl(
\Box\phi^I\Box\phi^J\Box\phi^K
-3\Box\phi^{I}\nabla_\alpha\nabla_\beta\phi^J\nabla^\alpha\nabla^\beta\phi^{K}
\nonumber\\&&
+2\nabla_\alpha\nabla_\beta\phi^I\nabla^\beta\nabla^\lambda\phi^J\nabla_\lambda\nabla^\alpha\phi^K
\bigr)\nabla_\mu\phi^L\nabla_\nu\phi^M
+g_{\mu\nu}\biggl[
2\nabla_\alpha G_{5I}\nabla_\beta\phi^IR^{\alpha\beta}
\nonumber\\&&
-\frac{1}{3}G_{5IJK}\left(
\Box\phi^I\Box\phi^J\Box\phi^K
-3\Box\phi^{I}\nabla_\alpha\nabla_\beta\phi^J\nabla^\alpha\nabla^\beta\phi^{K}
+2\nabla_\alpha\nabla_\beta\phi^I\nabla^\beta\nabla^\lambda\phi^J\nabla_\lambda\nabla^\alpha\phi^K
\right)
\nonumber\\&&
+G_{5IJK}\left(R_{\alpha\beta }\nabla^\alpha\phi^I\nabla^\beta\phi^J\Box\phi^K
-R_{\alpha\lambda\beta\sigma}\nabla^\alpha\phi^I\nabla^\beta\phi^J\nabla^\lambda\nabla^\sigma\phi^K\right)
-G_{5I,J}\left(\Box\phi^I\Box\phi^J-\nabla_\alpha\nabla_\beta\phi^I\nabla^\alpha\nabla^\beta\phi^J\right)
\nonumber\\&&
-\frac{1}{2}\nabla_\lambda G_{5IJK}
\left(\Box\phi^I\Box\phi^J-\nabla_\alpha\nabla_\beta\phi^I\nabla^\alpha\nabla^\beta\phi^J\right)
\nabla^\lambda\phi^K
-\nabla_\alpha G_{5I,J}\left(\Box\phi^I\nabla^\alpha\phi^J-\nabla_\beta\nabla^\alpha\phi^I\nabla^\beta\phi^J
\right)
\nonumber\\&&
+\nabla_\alpha G_{5IJK}\left(\Box\phi^I\nabla^\alpha\nabla^\lambda\phi^J
-\nabla^\alpha\nabla_\beta\phi^I\nabla^\beta\nabla^\lambda\phi^J\right)\nabla_\lambda\phi^K\biggr],
\end{eqnarray}
\begin{eqnarray}
J^\mu_I &=& - G_{2,\langle IJ\rangle}\nabla^\mu\phi^J
+G_{3IJK}\left(\Box\phi^J\nabla^\mu\phi^K+\nabla^\mu X^{JK}\right)
+2G_{3(I,J)}\nabla^\mu\phi^J
+2G_{4IJ}G^{\mu\nu}\nabla_\nu\phi^J
\nonumber\\&&
-2G_{4IJKL}\left(\Box\phi^J\nabla^\mu X^{KL}-\nabla^\mu\nabla^\nu\phi^J\nabla_\nu X^{KL}\right)
-G_{4IJKL}\nabla^{\mu}\phi^J\left(\Box\phi^K\Box\phi^L-\nabla_\alpha\nabla_\beta\phi^K
\nabla^\alpha\nabla^\beta\phi^L\right)
\nonumber\\&&
-2G_{4IJ,K}\left(\Box\phi^J\nabla^\mu\phi^K-\nabla^\mu\nabla^\nu\phi^J\nabla_\nu\phi^K\right)
-2G_{5(I,J)}G^{\mu\nu}\nabla_\nu\phi^J
-G_{5IJK}\biggl[
G^{\alpha\beta}\nabla_\alpha\nabla_\beta\phi^J\nabla^\mu\phi^K
\nonumber\\&&
+G^{\mu\nu}\nabla_\nu X^{JK}
+R^{\mu\nu}\Box\phi^J\nabla_\nu\phi^K
-R_{\alpha\beta}\nabla^\mu\nabla^\alpha\phi^J\nabla^\beta\phi^K
+R^\mu_{~\,\alpha\beta\nu}\nabla^\alpha\nabla^\beta\phi^J\nabla^\nu\phi^K
\biggr]
\nonumber\\&&
+G_{5IJK,L}\left[
\frac{1}{2}\nabla^\mu\phi^L\left(\Box\phi^J\Box\phi^K-\nabla_\alpha\nabla_\beta\phi^J
\nabla^\alpha\nabla^\beta\phi^K\right)
+\Box\phi^J\nabla^\mu X^{KL}-\nabla^\mu\nabla^\nu\phi^J\nabla_\nu X^{KL}
\right]
\nonumber\\&&
+G_{5IJKLM}\left[
\frac{1}{2}\nabla^\mu X^{JK}
\left(\Box\phi^L\Box\phi^M-\nabla_\alpha\nabla_\beta\phi^L
\nabla^\alpha\nabla^\beta\phi^M\right)
-\nabla_\nu X^{JK}\left(
\Box\phi^L\nabla^\mu\nabla^\nu\phi^M-\nabla^\lambda\nabla^\mu\phi^L
\nabla_\lambda\nabla^\nu\phi^M
\right)\right]
\nonumber\\&&
+\frac{1}{6}G_{5IJKLM}
\left(
\Box\phi^J\Box\phi^K\Box\phi^L
-3\Box\phi^{J}\nabla_\alpha\nabla_\beta\phi^K\nabla^\alpha\nabla^\beta\phi^L
+2\nabla_\alpha\nabla_\beta\phi^J\nabla^\beta\nabla^\lambda\phi^K\nabla_\lambda\nabla^\alpha\phi^L
\right)\nabla^\mu\phi^M,
\end{eqnarray}
and
\begin{eqnarray}
P_I&=&
G_{2, I}+\nabla_\mu G_{3J, I}\nabla^\mu\phi^J
+G_{4, I}R+
G_{4JK, I}
\left(\Box\phi^J\Box\phi^K-\nabla_\alpha\nabla_\beta\phi^J
\nabla^\alpha\nabla^\beta\phi^K\right)
-\nabla_\mu G_{5J, I}\nabla_\nu\phi^JG^{\mu\nu}
\nonumber\\&&
-\frac{1}{6}G_{5JKL, I}
\left(
\Box\phi^J\Box\phi^K\Box\phi^L
-3\Box\phi^{J}\nabla_\alpha\nabla_\beta\phi^K\nabla^\alpha\nabla^\beta\phi^L
+2\nabla_\alpha\nabla_\beta\phi^J\nabla^\beta\nabla^\lambda\phi^K\nabla_\lambda\nabla^\alpha\phi^L
\right).
\end{eqnarray}
The covariant equations of motion are thus given by
\begin{eqnarray}
{\cal G}_{\mu\nu}=0,\quad
\nabla_\mu J^\mu_I=P_I.
\end{eqnarray}
Note that $J_I^0\neq{\cal J}_I$
and $P_I\neq\partial{\cal P}/\partial\phi^I$,
where ${\cal J}_I$ and ${\cal P}$ are the background quantities used
in the main text; the equations of motion for the scalar fields
can be expressed in different ways by moving some part of
the left-hand side to the right-hand side.




\begin{thebibliography}{99}
%

\bibitem{Bennett:2012zja} 
  C.~L.~Bennett {\it et al.}  [WMAP Collaboration],
  arXiv:1212.5225 [astro-ph.CO].

\bibitem{Hinshaw:2012aka} 
  G.~Hinshaw {\it et al.}  [WMAP Collaboration],
  arXiv:1212.5226 [astro-ph.CO].

\bibitem{Ade:2013ktc} 
  P.~A.~R.~Ade {\it et al.}  [Planck Collaboration],
  arXiv:1303.5062 [astro-ph.CO].

\bibitem{Ade:2013zuv} 
  P.~A.~R.~Ade {\it et al.}  [Planck Collaboration],
  arXiv:1303.5076 [astro-ph.CO].

\bibitem{Ade:2013uln} 
  P.~A.~R.~Ade {\it et al.}  [Planck Collaboration],
  arXiv:1303.5082 [astro-ph.CO].

\bibitem{Kobayashi:2011nu} 
  T.~Kobayashi, M.~Yamaguchi and J.~Yokoyama,
  Prog.\ Theor.\ Phys.\  {\bf 126}, 511 (2011)
  [arXiv:1105.5723 [hep-th]].

\bibitem{Horndeski}
G.~W.~Horndeski,
Int.\ J.\ Theor.\ Phys.\ 10 (1974) 363-384.

\bibitem{Charmousis:2011bf} 
  C.~Charmousis, E.~J.~Copeland, A.~Padilla and P.~M.~Saffin,
  Phys.\ Rev.\ Lett.\  {\bf 108}, 051101 (2012)
  [arXiv:1106.2000 [hep-th]].

\bibitem{Kobayashi:2010cm} 
  T.~Kobayashi, M.~Yamaguchi and J.~Yokoyama,
  Phys.\ Rev.\ Lett.\  {\bf 105}, 231302 (2010)
  [arXiv:1008.0603 [hep-th]].

\bibitem{Deffayet:2010qz} 
  C.~Deffayet, O.~Pujolas, I.~Sawicki and A.~Vikman,
  JCAP {\bf 1010}, 026 (2010)
  [arXiv:1008.0048 [hep-th]].

\bibitem{Kamada:2010qe} 
  K.~Kamada, T.~Kobayashi, M.~Yamaguchi and J.~Yokoyama,
  Phys.\ Rev.\ D {\bf 83}, 083515 (2011)
  [arXiv:1012.4238 [astro-ph.CO]];
  K.~Kamada, T.~Kobayashi, T.~Takahashi, M.~Yamaguchi and J.~Yokoyama,
  Phys.\ Rev.\ D {\bf 86}, 023504 (2012)
  [arXiv:1203.4059 [hep-ph]].

\bibitem{Mizuno:2010ag} 
  S.~Mizuno and K.~Koyama,
  Phys.\ Rev.\ D {\bf 82}, 103518 (2010)
  [arXiv:1009.0677 [hep-th]].

\bibitem{Naruko:2011zk} 
  A.~Naruko and M.~Sasaki,
  Class.\ Quant.\ Grav.\  {\bf 28}, 072001 (2011)
  [arXiv:1101.3180 [astro-ph.CO]].

\bibitem{Kobayashi:2011pc} 
  T.~Kobayashi, M.~Yamaguchi and J.~Yokoyama,
  Phys.\ Rev.\ D {\bf 83}, 103524 (2011)
  [arXiv:1103.1740 [hep-th]].

\bibitem{Gao:2011mz} 
  X.~Gao,
  JCAP {\bf 1110}, 021 (2011)
  [arXiv:1106.0292 [astro-ph.CO]].

\bibitem{Gao:2011qe} 
  X.~Gao and D.~A.~Steer,
  JCAP {\bf 1112}, 019 (2011)
  [arXiv:1107.2642 [astro-ph.CO]].
 
\bibitem{DeFelice:2011uc} 
  A.~De Felice and S.~Tsujikawa,
  Phys.\ Rev.\ D {\bf 84}, 083504 (2011)
  [arXiv:1107.3917 [gr-qc]].

\bibitem{RenauxPetel:2011sb} 
  S.~Renaux-Petel,
  JCAP {\bf 1202}, 020 (2012)
  [arXiv:1107.5020 [astro-ph.CO]].
\bibitem{Gao:2011vs} 
  X.~Gao, T.~Kobayashi, M.~Yamaguchi and J.~Yokoyama,
  Phys.\ Rev.\ Lett.\  {\bf 107}, 211301 (2011)
  [arXiv:1108.3513 [astro-ph.CO]].

\bibitem{Gao:2012ib} 
  X.~Gao, T.~Kobayashi, M.~Shiraishi, M.~Yamaguchi, J.~Yokoyama and S.~Yokoyama,
  PTEP {\bf 2013}, 053E03 (2013)
  [arXiv:1207.0588 [astro-ph.CO]].

\bibitem{Takamizu:2013gy} 
  Y.~-i.~Takamizu and T.~Kobayashi,
  PTEP {\bf 2013}, no. 6, 063E03 (2013)
  [arXiv:1301.2370 [gr-qc]].
  
\bibitem{Frusciante:2013haa} 
  N.~Frusciante, S.~-Y.~Zhou and T.~P.~Sotiriou,
  JCAP {\bf 1307}, 020 (2013)
  [arXiv:1303.6628 [astro-ph.CO]].
  
\bibitem{DeFelice:2013ar} 
  A.~De Felice and S.~Tsujikawa,
  JCAP {\bf 1303}, 030 (2013)
  [arXiv:1301.5721 [hep-th]].

\bibitem{Tsujikawa:2013ila} 
  S.~Tsujikawa, J.~Ohashi, S.~Kuroyanagi and A.~De Felice,
  Phys.\  Rev.\ D {\bf 88}, 023529 (2013)
  [arXiv:1305.3044 [astro-ph.CO]].

\bibitem{Easson:2007dh} 
  D.~A.~Easson, R.~Gregory, D.~F.~Mota, G.~Tasinato and I.~Zavala,
  JCAP {\bf 0802}, 010 (2008)
  [arXiv:0709.2666 [hep-th]].

\bibitem{Huang:2007hh} 
  M.~-x.~Huang, G.~Shiu and B.~Underwood,
  Phys.\ Rev.\ D {\bf 77}, 023511 (2008)
  [arXiv:0709.3299 [hep-th]].

\bibitem{Langlois:2008mn} 
  D.~Langlois and S.~Renaux-Petel,
  JCAP {\bf 0804}, 017 (2008)
  [arXiv:0801.1085 [hep-th]].

\bibitem{Langlois:2008wt} 
  D.~Langlois, S.~Renaux-Petel, D.~A.~Steer and T.~Tanaka,
  Phys.\ Rev.\ Lett.\  {\bf 101}, 061301 (2008)
  [arXiv:0804.3139 [hep-th]].

\bibitem{Langlois:2008qf} 
  D.~Langlois, S.~Renaux-Petel, D.~A.~Steer and T.~Tanaka,
  Phys.\ Rev.\ D {\bf 78}, 063523 (2008)
  [arXiv:0806.0336 [hep-th]].


\bibitem{Langlois:2009ej} 
  D.~Langlois, S.~Renaux-Petel and D.~A.~Steer,
  JCAP {\bf 0904}, 021 (2009)
  [arXiv:0902.2941 [hep-th]].
\bibitem{RenauxPetel:2009sj} 
  S.~Renaux-Petel,
  JCAP {\bf 0910}, 012 (2009)
  [arXiv:0907.2476 [hep-th]].


\bibitem{Padilla} 
  A.~Padilla and V.~Sivanesan,
  JHEP {\bf 1304}, 032 (2013)
  [arXiv:1210.4026 [gr-qc]].


\bibitem{Deffayet:2013lga} 
  C.~Deffayet and D.~ASteer,
  arXiv:1307.2450 [hep-th].

\bibitem{Nicolis:2008in} 
  A.~Nicolis, R.~Rattazzi and E.~Trincherini,
  Phys.\ Rev.\ D {\bf 79}, 064036 (2009)
  [arXiv:0811.2197 [hep-th]].

\bibitem{Deffayet:2009wt} 
  C.~Deffayet, G.~Esposito-Farese and A.~Vikman,
  Phys.\ Rev.\ D {\bf 79}, 084003 (2009)
  [arXiv:0901.1314 [hep-th]].
  
\bibitem{Deffayet:2011gz} 
  C.~Deffayet, X.~Gao, D.~A.~Steer and G.~Zahariade,
  Phys.\ Rev.\ D {\bf 84}, 064039 (2011)
  [arXiv:1103.3260 [hep-th]].


\bibitem{disf}
D.~Bettoni and S.~Liberati,
  arXiv:1306.6724 [gr-qc];
  M.~Zumalacarregui and J.~Garcia-Bellido,
  arXiv:1308.4685 [gr-qc].

\bibitem{deRham:2011by} 
  C.~de Rham and L.~Heisenberg,
  Phys.\ Rev.\ D {\bf 84}, 043503 (2011)
  [arXiv:1106.3312 [hep-th]].

\bibitem{Narikawa:2013pjr} 
  T.~Narikawa, T.~Kobayashi, D.~Yamauchi and R.~Saito,
  Phys.\  Rev.\ D {\bf 87}, 124006 (2013)
  [arXiv:1302.2311 [astro-ph.CO]].


\bibitem{Deffayet:2010zh}
  C.~Deffayet, S.~Deser and G.~Esposito-Farese,
  Phys.\ Rev.\ D {\bf 82}, 061501 (2010)
  [arXiv:1007.5278 [gr-qc]].

\bibitem{Bi-galileon} 
  A.~Padilla, P.~M.~Saffin and S.~-Y.~Zhou,
  JHEP {\bf 1012}, 031 (2010)
  [arXiv:1007.5424 [hep-th]];
  A.~Padilla, P.~M.~Saffin and S.~-Y.~Zhou,
  JHEP {\bf 1101}, 099 (2011)
  [arXiv:1008.3312 [hep-th]].



\bibitem{Sivanesan:2013tba}
  V.~Sivanesan,
  arXiv:1307.8081 [gr-qc].


\bibitem{Gordon:2000hv} 
  C.~Gordon, D.~Wands, B.~A.~Bassett and R.~Maartens,
  Phys.\ Rev.\ D {\bf 63}, 023506 (2001)
  [astro-ph/0009131].

\bibitem{Starobinsky:1994mh} 
  A.~A.~Starobinsky and J.~Yokoyama,
  gr-qc/9502002.

\bibitem{deRham:2010eu} 
  C.~de Rham and A.~J.~Tolley,
  JCAP {\bf 1005}, 015 (2010)
  [arXiv:1003.5917 [hep-th]].
  
\bibitem{Goon:2011uw} 
  G.~Goon, K.~Hinterbichler and M.~Trodden,
  Phys.\ Rev.\ Lett.\  {\bf 106}, 231102 (2011)
  [arXiv:1103.6029 [hep-th]].
\bibitem{Hinterbichler:2010xn} 
  K.~Hinterbichler, M.~Trodden and D.~Wesley,
  Phys.\ Rev.\ D {\bf 82}, 124018 (2010)
  [arXiv:1008.1305 [hep-th]].

\bibitem{RenauxPetel:2011dv} 
  S.~Renaux-Petel,
  Class.\ Quant.\ Grav.\  {\bf 28}, 182001 (2011)
  [Erratum-ibid.\  {\bf 28}, 249601 (2011)]
  [arXiv:1105.6366 [astro-ph.CO]].

\bibitem{RenauxPetel:2011uk} 
  S.~Renaux-Petel, S.~Mizuno and K.~Koyama,
  JCAP {\bf 1111}, 042 (2011)
  [arXiv:1108.0305 [astro-ph.CO]].

\bibitem{Koyama:2013wma} 
  K.~Koyama, G.~W.~Pettinari, S.~Mizuno and C.~Fidler,
  arXiv:1303.2125 [astro-ph.CO].

\bibitem{eftinf}
  C.~Cheung, P.~Creminelli, A.~L.~Fitzpatrick, J.~Kaplan and L.~Senatore,
  JHEP {\bf 0803}, 014 (2008)
  [arXiv:0709.0293 [hep-th]];
  S.~Weinberg,
  Phys.\ Rev.\ D {\bf 77}, 123541 (2008)
  [arXiv:0804.4291 [hep-th]].
 
\bibitem{multieft} 
  L.~Senatore and M.~Zaldarriaga,
  JHEP {\bf 1204}, 024 (2012)  [arXiv:1009.2093 [hep-th]];  
  T.~Noumi, M.~Yamaguchi and D.~Yokoyama,
  JHEP {\bf 1306}, 051 (2013)  [arXiv:1211.1624 [hep-th]].  

\bibitem{Gleyzes:2013ooa} 
  J.~Gleyzes, D.~Langlois, F.~Piazza and F.~Vernizzi,
  JCAP {\bf 1308}, 025 (2013)  [arXiv:1304.4840 [hep-th]].  


\end{thebibliography}
\end{document}